\documentclass[preprint,12pt]{elsarticle}

\usepackage{amssymb}
\usepackage{amsmath}

\usepackage{hyperref}
\usepackage{multirow}
\usepackage{siunitx}
\usepackage{doi}
\usepackage{xcolor}
\journal{Neurocomputing}

\usepackage{array}
\newcolumntype{M}[1]{>{\centering\arraybackslash}m{#1}}
\usepackage{seqsplit}

\begin{document}

\begin{frontmatter}

\title{Real-time closed-loop protocol to assess neural variability in temporal coding}

\author{Alberto Ayala\corref{cor1}}
\ead{alberto.ayala@estudiante.uam.es}
\author{Angel Lareo}
\ead{angel.lareo@uam.es}
\author{Pablo Varona}
\ead{pablo.varona@uam.es}
\author{Francisco B. Rodriguez\corref{cor1}}
\ead{f.rodriguez@uam.es}

\cortext[cor1]{Corresponding author.}

\affiliation{organization={Grupo de Neurocomputación Biológica, Departamento de Ingeniería Informática, Escuela Politécnica Superior, Universidad Autónoma de Madrid},
	addressline={C. Francisco Tomás y Valiente, 11}, 
	city={Madrid},
	postcode={28049}, 
	country={Spain}}

\begin{abstract}

Understanding temporal coding in neural systems is essential for decoding brain communication and advancing knowledge of neural information processing. Neural activity often conveys information through spike sequences with stereotypical temporal structures linked to specific functions. However, these sequences are subject to variability introduced by neural dynamics. Real-time closed-loop stimulation is a powerful approach to study temporal coding through adaptive control. In this work, we evaluate how a closed-loop protocol adapts to this variability to drive neural dynamics toward a desired state. It computes the Victor-Purpura distance to quantify similarity between spike sequences generated by the neural system and a triggering pattern. If the protocol determines that a neural sequence is similar to the trigger pattern, it applies stimulation to the system. This allows for an analysis of whether the system’s responses are consistent and facilitates the identification of varying spike sequences that can be considered instances of the same functional temporal code. We designed two validation experiments using the Hindmarsh-Rose model: (i) detection of a temporal code and delivery of stimulation to produce brief interspersed bursts, and (ii) detection of burst onset in chaotic activity followed by inhibitory stimulation to regularize it. Gaussian noise was progressively injected to increase variability. The protocol exhibited high degree of adaptability to variability and was effective in achieving the target dynamics. The results reported in this paper suggest that adaptive closed-loop stimulation can enhance experimental methodologies for studying neural coding under realistic variability conditions.

\end{abstract}

\begin{highlights}
\item We evaluate a real-time closed-loop stimulation protocol under variable activity.
\item The protocol drives neural dynamics toward desired states despite variability.
\item Demonstrates superior adaptability compared to open-loop stimulation.
\item Meets strict real-time requirements for experimental applications.
\end{highlights}


\end{frontmatter}
\section{Introduction}\label{Introduction}

How neural systems encode information is one of the main areas of research in contemporary neuroscience. Several coding schemes have been proposed to explain this process \cite{nadspike}, but it remains a challenging problem because information is encoded across multiple time scales, in multi-scale systems, and often in a multiplexed manner \cite{baker_multiplexed_2013,panzeri_sensory_2010}. 

\begin{figure}
	\centering
	\includegraphics[width=\textwidth]{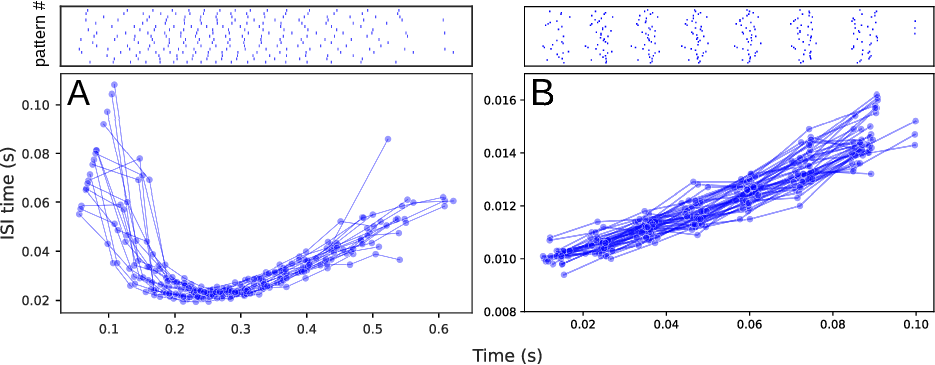}
	\caption{Examples of variable families of spiking patterns in biological and model neural systems. Panel A shows the activity patterns emitted by a weakly electric fish \cite{Lareo2016, Lareo2017} that convey the same behavioral message despite natural variability in spike timing. These patterns correspond to acceleration signals associated with agonistic behaviors in mormyrid fish. Panel B shows an artificial family of activity spike sequences produced by the Hindmarsh-Rose model under the same underlying dynamic regime and exhibiting  variability similar to that observed in biological systems. In the upper part of both panels, a raster plot of the activity patterns is displayed. It is important to note that spike sequences are aligned to the first spike.}

\label{Fig:target-code-e-fish}
\end{figure}

In this work, we focus on the temporal coding scheme \cite{rodriguez2004analysis,lestienne_spike_2001}, which posits that the precise timing of spikes in neural activity sequences carries relevant information in the coding process. Several studies have reported precise spatio-temporal spiking patterns in different areas of the brain that encode the underlying state of the system \cite{abeles_spatiotemporal_1993,nadasdy_replay_1999,cariani_survey_2025,panzeri_structures_2022,ma_neural_2014}. Temporal coding is also central to processing external stimuli in several sensorimotor systems, such as sound localization and communication \cite{sabourin_temporal_2010}, odor discrimination \cite{laurent_odor_2001,Laurent19963837}, visual stimulus encoding \cite{Victor1996}, and electroreception \cite{Carlson2004,baker_multiplexed_2013,lareo2022modeling}. It is important to mention that these temporal sequences are generated with a stereotypical temporal structure (representing a neural code), but with intrinsic variability in both the number of action potentials and the temporal intervals between them \cite{rieke1999spikes}. In this work, a \textit{code} is defined as a temporal sequence of discrete events (neural spikes) characterized by the relative structure of their occurrence times. As an example, Panel A of Figure \ref{Fig:target-code-e-fish} shows an activity code emitted by a weakly electric fish. This fish generates sequences of electric field potentials in the water to explore the environment and communicate with other members of the same species. These sequences correspond directly to the output of the electromotor system \cite{lareo2022modeling} and they translate into behaviors such as aggression \cite{kramer_attack_1976,Carlson2004,carlson_electric_2002}, which underlines their functional relevance and the need to study their mechanisms. In addition, Panel B shows an example of activity patterns produced by a neural model that exhibit variability similar to that observed in biological systems.

Most neural systems operate in a history-dependent manner. Although these systems have been traditionally studied from the perspective of how they encode information through unidirectional stimulation, current closed-loop approaches complement this perspective. They allow real-time exploration of neural dynamics and activity-dependent stimulation, making it possible to study complex neural dynamics that are difficult to assess with traditional methods and to relate the system's response to its internal state. 

In this manuscript, we study the degree of adaptability of an activity-dependent stimulation protocol to the variability with which neural codes are produced. This protocol is based on the Temporal Code-Driven Stimulation (TCDS) method \cite{Lareo2016,Lareo2017,ayala2021closed,ayala_neural_2025,10.1007/978-3-031-34107-6_43}, which detects spike sequences in real-time in the activity of a neural system with a predefined temporal structure. Upon detection, a predefined stimulus is delivered. This protocol uses the Victor-Purpura distance \cite{Victor1996} to establish whether the neural codes produced by the system, despite exhibiting spike-time jitter, retain a temporal structure similar to that of a triggering code. In this way, the protocol can detect neural codes while accounting for the variability with which they are produced and deliver stimulation after detection. In contrast to other closed-loop approaches, this protocol uses the similarity between temporal codes (rather than variables such as firing rate or signal amplitude) to guide real-time stimulation. We hypothesize that if the system responds comparably, the neural code that triggers stimulation may convey the same information, that is, the detected spike sequences are potentially functionally equivalent.

Neural systems encode information at the millisecond scale or below, so the protocol must operate on this timescale. To comply with these strict temporal requirements, the protocol has been implemented for the Real-Time eXperiment Interface (RTXI) tool \cite{patel_hard_2017}. This open-source framework has been specifically designed for real-time biological experimentation. In addition to being appropriate for the context of this work, it is widely used by other laboratories \cite{skocik_real-time_2013,boyle_eeg_2013}, which facilitates the dissemination of this protocol and enables its application to a wide variety of experiments.

In this paper, we study how the protocol adapts to neural variability to drive system activity toward a desired dynamic state. Two sets of experiments were performed in which the case study was the Hindmarsh-Rose (HR) neural model \cite{Hindmarsh1984} configured to produce two types of bursting activity. In addition, variability was induced in its activity by injecting Gaussian noise into the model, progressively increasing its level during the experiments. In the first set of experiments, the HR model was set up to produce regular bursting activity. The protocol was configured to detect these bursts, and to deliver stimulation to elicit brief bursts interspersed between them. In the second experiment, the protocol regularizes the chaotic activity of the HR model. An example of this neural code, along with the variability with which it was produced by the model, is shown in the right panel of Figure \ref{Fig:target-code-e-fish}. These experiments were performed in both real-time and simulated environments. In addition, it was verified whether it was possible to condition the model's activity through unidirectional (open-loop) stimulation, i.e., without taking into account the activity prior to stimulation.

This work builds on the line opened by Ayala et al. \cite{ayala_2025} and addresses two aspects that remained unexplored in that study. First, we designed and implemented a new experiment in which the protocol was used to regularize chaotic activity. Second, we evaluated the protocol in a real-time environment to assess its adaptability under realistic conditions. Specifically, we assess whether the protocol can deliver stimulation after detecting the triggering neural code and consequently drive the model's activity toward the desired dynamic state in a robust manner. The protocol's ability to condition the model's activity is compared with that of open-loop stimulation. Furthermore, we investigate whether the results obtained in the real-time experiments are consistent with those observed in the simulated environment. Finally, the latency of the real-time experiments is analyzed to evaluate the suitability of the protocol for studying temporal coding in neural systems through activity-dependent stimulation that accounts for intrinsic variability.

\section{Methods}

This section describes the methodology used to evaluate the adaptability of a real-time closed-loop stimulation protocol under variable neural activity. First, we detail the real-time operation of the Temporal Code-Driven Stimulation (TCDS) algorithm and its use of the Victor-Purpura distance for spike train comparison (Subsection \ref{protocol-description}). Next, we explain how variability was systematically produced in the Hindmarsh-Rose neural model to emulate realistic conditions (Subsection \ref{characterization-of-variability-induced-hr-model}). Finally, we present two experimental paradigms in which the protocol was applied to modulate the Hindmarsh-Rose  model’s activity, assessing its ability to maintain performance under increasing variability (Subsection \ref{experimental-design}).

\subsection{Real-time implementation of the TCDS protocol and the Victor–Purpura distance} \label{protocol-description}

\begin{figure}
	\includegraphics[width=\textwidth]{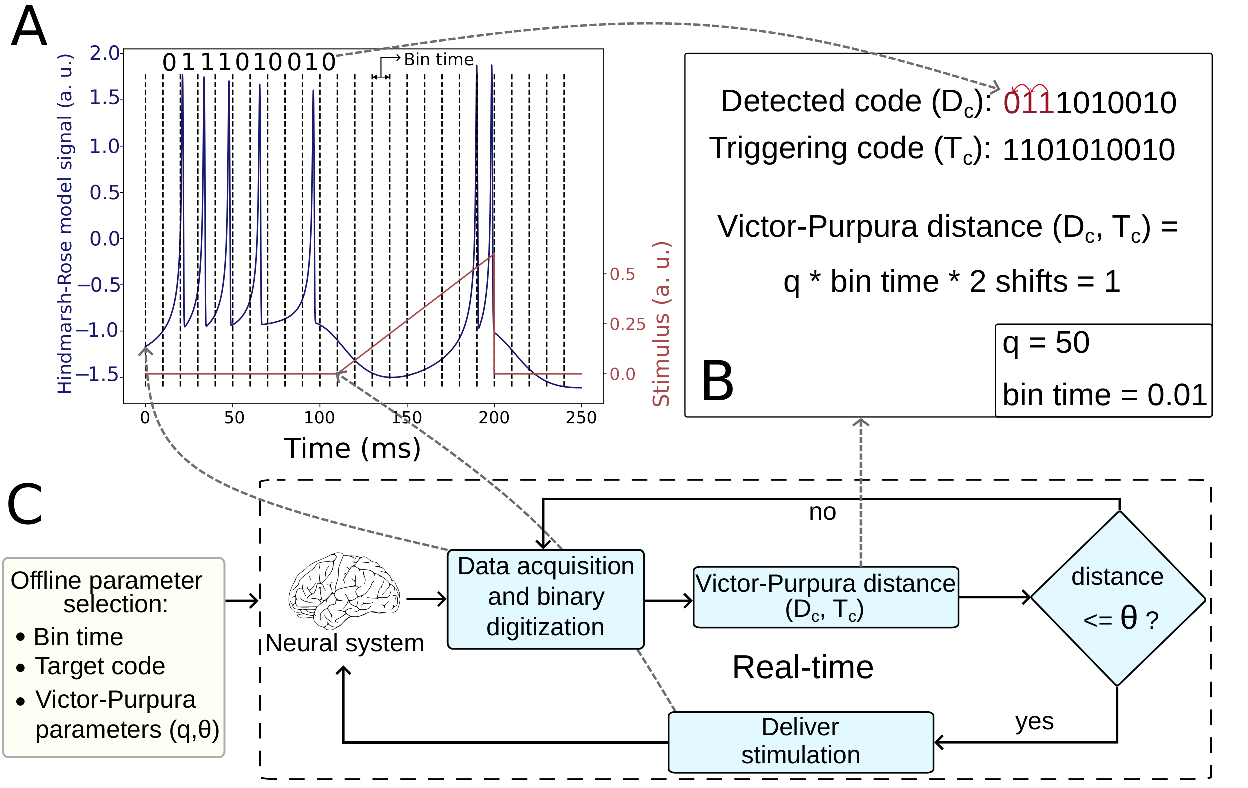}\hfill
	\caption{Flowchart of the proposed protocol. Panel A illustrates data acquisition and the binary digitization of Hindmarsh–Rose model activity. Panel B explains the Victor-Purpura distance computation. The distance between the detected code shown in Panel A ($D_{c}$) '0111010010' and the triggering code ($T_{C}$) '1101010010' is computed by transforming one code into the other through two bit shifts of 10 ms (bin time). Thus, the distance is 1 ($q$ = 50). If this result is smaller than or equal to the protocol threshold parameter ($\theta$), stimulation is triggered. Panel C shows the offline and online tasks of the protocol. The parameters are set offline, at the beginning of the experiment. The dashed box illustrates the flow sequence of the operations executed during each real-time task, which are performed at a specific frequency (10 kHz = 10000 times per second).}\label{ProtocolSchema}
\end{figure}

As outlined in the Introduction (Section \ref{Introduction}), the foundation of the protocol is the Temporal Code-Driven Stimulation (TCDS) algorithm \cite{Lareo2016,Lareo2017,ayala2021closed,ayala2022closed}. As shown in Figure \ref{ProtocolSchema}, TCDS acquires a biological system signal in real-time and segments it into consecutive time windows of predefined length (bin time protocol parameter). Each segment is referred to as a bit. When a spike occurs within a given time window, the bit is assigned the value 1; otherwise, it is assigned 0 (see Figure \ref{ProtocolSchema}, Panel A). In this work, a code is defined as a sequence of bits of fixed length that represents the temporal structure of a sequence of spikes discretized by the bin time parameter. It is important to note that the protocol resolution is determined by the bin time. Therefore, the optimal bin time captures spike events without loss of information. In practice, this parameter is selected based on the characteristic time scale of the analyzed activity. For every code detected in real-time, the Victor-Purpura distance \cite{Victor1996} with respect to a triggering neural code is computed. This metric has been used to study neural encoding by analyzing responses to stimuli, supporting tasks such as connectivity estimation \cite{tezuka_kernel_2017} and tactile classification \cite{soares_victor-purpura_2025}. It was also used to analyze the clustering of auditory responses \cite{dimitrov_inhibition_2014} and evaluate the repeatability and linearity of sensory signals \cite{yi_spike_2019}.

The Victor-Purpura metric defines the distance as the minimal cost required to transform one spike train into another through three elementary operations: i) inserting a spike with a fixed cost of 1, ii) deleting a spike with a fixed cost of 1, iii) shifting an existing pulse from time $t_1$ to time $t_2$. The cost of this last operation is $q \cdot |t_1 - t_2|$, where $q$ ($s^{-1}$) determines the cost of shifting one spike in time. Therefore, if the value of $q$ is 0, shifting a spike has no cost and the distance becomes only the difference in the number of spikes between both trains (rate coding scheme). Conversely, if the value of $q$ is very high, shifting a spike costs more than deleting and inserting, making the metric extremely sensitive to the synchrony between compared trains. Thus, $q$ controls sensitivity to temporal misalignment. The Victor-Purpura metric provides a measure of similarity between the temporal structure of two codes. Whenever the distance between the detected code and the triggering code falls below a predefined threshold ($\theta$), the protocol treats them as temporally similar and immediately delivers a predefined stimulus to the system in real-time. This threshold defines the protocol's tolerance for temporal variability in the structure of the detected code, enabling its identification under non-deterministic conditions. Analyzing whether the system's response to stimulation triggered by these codes is comparable allows the study of potentially equivalent neural codes. Thus, consistent responses show that variable spike sequences can act as the same underlying code.

\subsection{Case study: Hindmarsh–Rose neural model with injected noise}\label{characterization-of-variability-induced-hr-model}

We tested the performance of a closed-loop stimulation protocol for adapting to the noise‑induced variability of the Hindmarsh–Rose (HR) neural model \cite{Hindmarsh1984} and for conditioning its activity to a target dynamical state. This model replicates both spiking and bursting activity types, present in the dynamics of individual neurons, e.g., in central pattern generators \cite{szucs2000interacting,amaducci2019rthybrid,reyes2020automatic} and its activity has been previously conditioned and studied \cite{emiroglu_nonlinear_2022} even using hardware implementations \cite{liu_periodic_2024} as in this work.

The dynamics of the model are defined by the following set of three differential equations \ref{HREq}:

\begin{equation} \label{HREq}
	\begin{aligned}
		& \dot{x}=y-a x^3+b x^2-z+I_{reg}+I_{noise}+I_{stim},\\
		& \dot{y}=c-d x^2-y,\\
		& \dot{z}=r\left(s\left(x-x_1\right)-z\right).
	\end{aligned}
\end{equation}

\noindent In the model, $x$ corresponds to the membrane potential, $y$ represents the recovery variable linked to fast ionic currents, and $z$ accounts for a slow ionic current adaptation. The HR model, first introduced by Hindmarsh and Rose \cite{Hindmarsh1984} and widely used in subsequent studies \cite{Barrio2017}, was parameterized with r = $0.0021$, a = $1$, b = $3$, c = $1$, d = $5$, s = $4$, and $x_{1}$ = $-8/5$. Under this configuration, the system can shift from bursting to spiking activity regime by adjusting the control parameter $I_{reg}$ \cite{innocenti_dynamics_2009, innocenti_dynamical_2007}. For the present work, two different dynamical regimes were considered: regular bursting at $I_{reg}$ = 1.9 and chaotic bursting at $I_{reg}$ = 3.281. As mentioned in the Introduction (Section \ref{Introduction}), biological activity naturally exhibits intrinsic variability. To account for this property, Gaussian noise was introduced as an external current ($I_{noise}$ in Eq. \ref{HREq}), ensuring variability in the simulated neural dynamics.  Lastly, $I_{stim}$ denotes the stimulation current applied in both closed-loop and open-loop sessions to drive the system toward the desired dynamical state. 
\subsection{Experimental design: brief-burst generation and regularization}\label{experimental-design}

This subsection details the design of two experiments used to study the protocol's adaptability to variability in neural activity. In both experiments, the closed-loop protocol searched for a triggering neural code and stimulated the Hindmarsh-Rose model with noise‑induced variability (described in Section \ref{characterization-of-variability-induced-hr-model}) to drive its activity toward a target dynamic state. Furthermore, in both scenarios the experiments were carried out in a simulated (software) and real (hardware-constrained). In the design of the first experiment, the protocol conditioned the model's activity to generate brief bursts interspersed within the regular bursts produced by the model. In the second experiment, the protocol regularized the chaotic activity of the model.

\subsubsection{Closed‑loop generation of brief bursts in the Hindmarsh-Rose model}\label{ExpBbs}

For this experiment, in the simulated environment, the HR model was configured as described in Section \ref{characterization-of-variability-induced-hr-model} to produce regular bursting activity ($I_{reg}$ = 1.9 in Eq. \ref{HREq}). In addition, Gaussian noise ($I_{noise}$ in Eq. \ref{HREq}) with zero mean and a gradually increasing standard deviation was added to the model. The noise level supplied is expressed as a percentage of the external current applied to the model. For example, when $I_{reg}$ equals 1.9, 25\% of this value corresponds to a noise standard deviation of 0.475.

The noise intensity was varied from 0\% to 26\%, and at each stage, 10 trials were performed to ensure robustness of the results. The closed-loop protocol was configured to detect complete bursts consistently, whose temporal structure becomes variable due to the added noise and to stimulate the model ($I_{stim}$ in Eq. \ref{HREq}) using a positive ramp after detection to generate brief bursts containing 2-3 spikes in its activity (parameter settings are presented in Table \ref{table:experimental-parameters}). To validate the closed-loop results, we evaluated whether the HR model could be conditioned with the same effectiveness in open-loop sessions. To this end, we recorded the stimulus starting times from the closed-loop session and subsequently delivered ramp stimuli at those same times, without conditioning stimulus delivery on the model’s ongoing activity. For each closed-loop session, a corresponding open-loop session was conducted. Several results obtained using this experimental design were previously reported by Ayala et al. \cite{ayala_2025}.

\begin{figure}
	\centering
	\includegraphics[width=\textwidth]{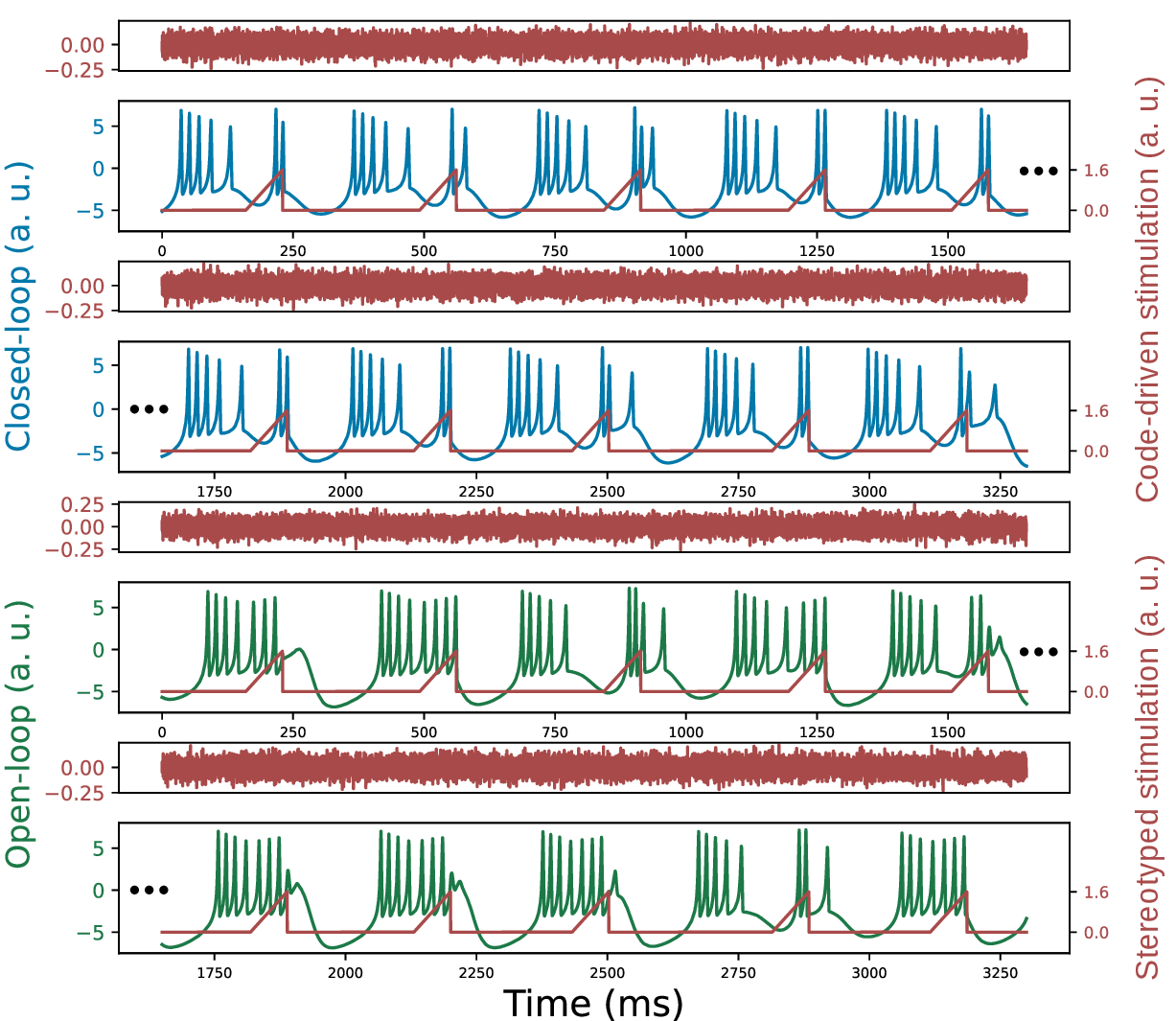}
	\caption{Activity of the Hindmarsh-Rose model with Gaussian noise of 0.06 standard deviation (comparable to a 3\% noise level) in real-time closed-loop and open-loop sessions. In both cases, ramp stimuli were delivered at the same time points. Closed-loop stimulation robustly elicited brief bursts (the desired dynamic state) up to 97.45\% of the stimuli. In contrast, open-loop stimulation elicited brief bursts in only 38.24\% of the stimuli.} \label{Fig:protocol-example-bbs-rt-0.06}
\end{figure}

This experiment was also applied in a hardware constrained real-time environment (described in Section \ref{real-time-env}). In this case, an electronic neuron (shown in Figure \ref{real-time-set-up}) implementing the HR model was used as a case study. The intensity ($I_{reg}$ in Eq. \ref{HREq}) of this device can be regulated manually  with a potentiometer to change the activity regime. Thus, although it was configured in a regular bursting regime, the parameters of the electronic neuron are not identical to those of the HR software model. Therefore, the protocol parameters were adjusted to produce brief bursts in the activity of the electronic neuron (parameter settings are presented in Table \ref{table:experimental-parameters}). In the real-time environment, three 5-minute experiments were carried out at different noise levels: no noise, 3\%, and 25\%. The results were compared with those obtained by Ayala et al. \cite{ayala_2025}. An example of these experiments is shown in Figure \ref{Fig:protocol-example-bbs-rt-0.06}. The two upper panels correspond to a segment of a closed-loop session with 3\% noise, where the protocol conditions the model to generate brief bursts. The two lower panels show the same time segment in the corresponding open-loop session, where despite delivering the stimuli to the model at the same time points, it is not possible to condition its activity with the same effectiveness.

The adaptability of the closed-loop and open-loop protocols was assessed by calculating the percentage of stimuli that elicited a brief burst, relative to the total number of stimuli delivered. The mean percentage and standard deviation were also computed, enabling an adaptability comparison. In addition, the codes emitted by the HR model prior to stimulation (pre-stimulation codes) were examined in both closed-loop and open loop sessions for each level of noise. This analysis verifies whether the protocol adapts to the variability of the neural code produced by the model and provides a comparison with the open-loop condition. Furthermore, as detailed in Ayala et al. \cite{ayala_neural_2025}, the activity of the HR model under stimulation delivered by both protocols was characterized. For a given experiment with a specific noise level, segments of HR model activity during stimulation were selected. From each segment, 15 recurrence metrics (laminarity, determinism, etc.) \cite{marwan_recurrence_2007} were extracted (using PyRQA software tool \cite{rawald_pyrqaconducting_2017}) based on its computed recurrence plot \cite{Eckmann1987}. A Principal Component Analysis (PCA) \cite{hotelling1992relations} was then applied, and the first two components were plotted, allowing a comparison of how the model activity was conditioned under closed-loop versus open-loop stimulation. Finally, for the set of recurrence metrics, the silhouette coefficient was calculated, providing a quantitative characterization of the differences in conditioning the HR model between the closed-loop and open-loop stimulation protocols.

\begin{table}
	\centering
	\begin{tabular}{M{0.4cm}|M{0.8cm}|M{2.1cm}|M{1.4cm}|M{0.6cm}|M{0.9cm}|M{1.7cm}|M{1.7cm}|} 
		\cline{2-8}
		 &$\mathbf{I_{reg}}$&\textbf{triggering code}&\textbf{bin}&\textbf{VP} $\mathbf{\theta}$&\textbf{VP}$\mathbf{q}$&\textbf{stimulus amplitude}&\textbf{stimulus duration}\\ [0.4ex]
		\hline
		\multicolumn{1}{|c|}{$\mathbf{BB}$}&1.9&\seqsplit{1101010010}& 10 a. u.& 1.5 &50&0.3&90 a. u.\\ 
		\hline
		\multicolumn{1}{|c|}{$\mathbf{BB_{rt}}$}& - &\seqsplit{1010101000100}& 10 ms& 1.5 &50&0.8&70 ms\\ 
		\hline
		\multicolumn{1}{|c|}{$\mathbf{R}$}&3.281&\seqsplit{10101001010100100100}& 5 ms& 3 &100&-0.054&100 ms\\ 
		\hline
	\end{tabular}
	\caption{Parameter settings for both experiments. BB experiments eliciting brief bursts were conducted in simulated ($BB$) and real-time ($BB_{rt}$) environments. $R$ indicates values used in the regularization experiment.}
	\label{table:experimental-parameters}
\end{table}

\subsubsection{Closed-loop regularization of chaotic activity in the Hindmarsh-Rose modely} \label{Exp-Reg}

In the simulated environment version of this experiment, the closed-loop protocol described in Section \ref{protocol-description} was applied to detect the onset of bursts produced by the HR model configured (as described in Section \ref{characterization-of-variability-induced-hr-model}) in a chaotic bursting regime ($I_{reg}$ = $3.281$ in Eq. \ref{HREq}). After detection, deliver an inhibitory stimulation to regularize its burst production (parameter settings are presented in Table \ref{table:experimental-parameters}). The target is to drive the chaotic model to produce bursts with a regular period, using the first spike of each burst as a reference. It is important to note that the regularization of the bursting activity of the Hindmarsh-Rose model has been previously studied using other approaches \cite{NEURIPS2023_d6db7eb6,emiroglu_nonlinear_2022}. As in the brief burst generation experiment, the adaptability of the protocol has been evaluated by supplying Gaussian noise with zero mean and gradually increasing standard deviation to the model. 

In this experiment, noise levels ranged from 0\% to 7\%, and for each level, 10 trials were performed to enable statistical analysis. Similarly to the brief burst generation experiment, open-loop stimulation sessions were conducted for an adaptability comparison.

In addition to software simulation, this experiment was also applied in a hardware-constrained real-time environment. In this case, a software-based implementation of the HR model was used for real-time execution. Three 5-minute experiments were carried out at different noise levels: no noise, 0.45\%, and 5\%. The results were compared with those obtained in the simulated environment.

\begin{figure}
	\centering
	\includegraphics[width=\textwidth]{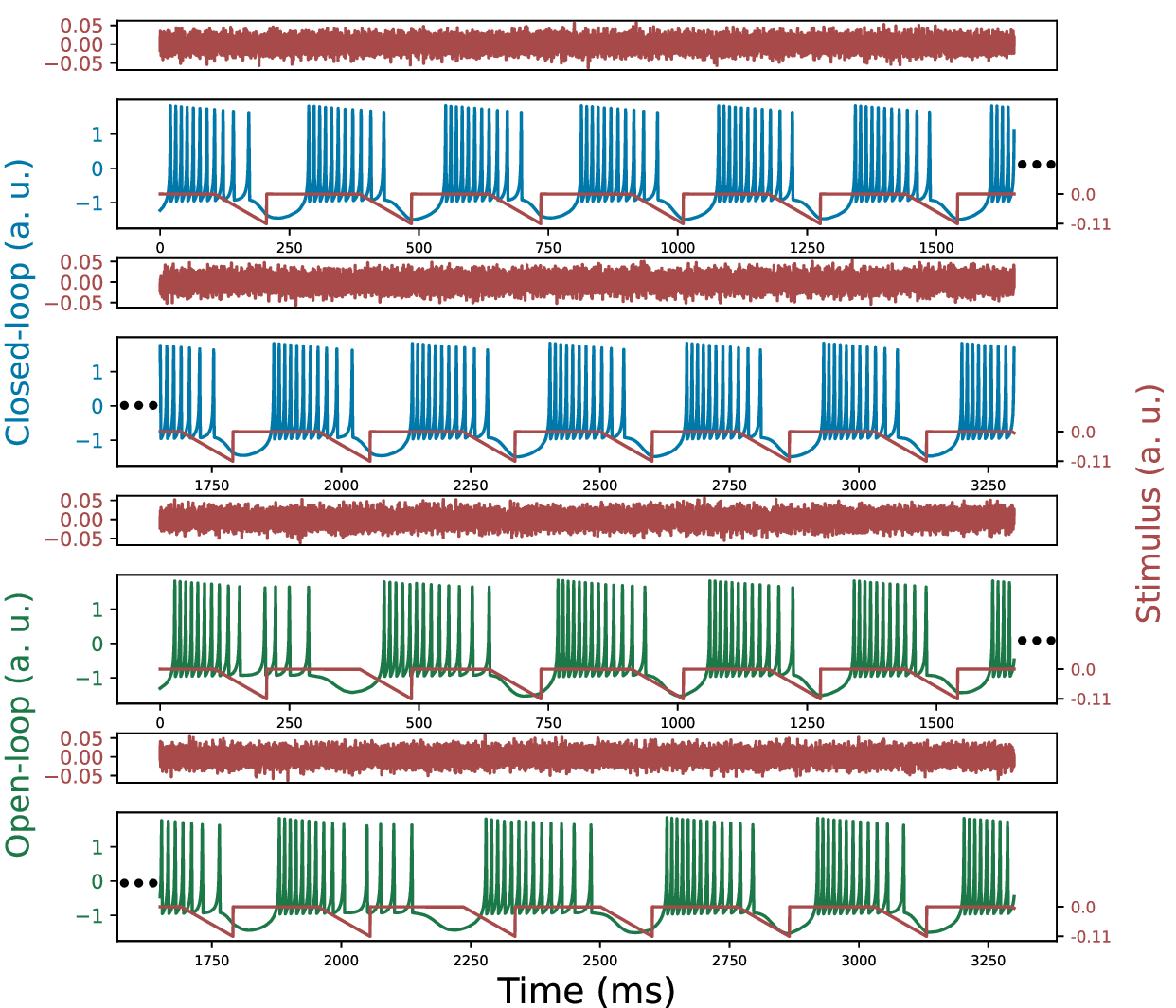}
	\caption{Activity of the Hindmarsh-Rose model with 0.45\% Gaussian noise level in real-time closed-loop and open-loop sessions. Closed-loop stimulation robustly regularizes the model activity (1.21 period standard deviation). In opposition, open-loop stimulation regularization is less consistent (20.73 period standard deviation).} \label{Fig:protocol-example-reg-rt-0.015}
\end{figure}

An example of this experiment is shown in Figure \ref{Fig:protocol-example-reg-rt-0.015}. The top two panels illustrate how the closed-loop protocol consistently regularizes the model’s burst production despite the presence of noise (0.45\%). In contrast, the bottom panels show that open-loop stimulation fails to achieve burst regularization with the same robustness.

The performance of both protocols was evaluated by measuring the mean and standard deviation of the burst period across trials. Activity prior to stimulation was also analyzed, following the same procedure as in Section \ref{ExpBbs}. Furthermore, as detailed in Ayala et al. \cite{ayala_neural_2025}, segments of activity (7.5 s) were selected, and recurrence plots and 15 associated metrics were computed. Finally, PCA was applied, and the first two components were plotted. This activity characterization is detailed in Figure \ref{Fig:cluster-reg-rt}. In addition, for the recurrence data, the silhouette coefficient was calculated to quantify the differences in the activity between both stimulation methodologies.

\subsection{Hardware-constrained real-time setup} \label{real-time-env}

\begin{figure}
	\centering
	\includegraphics[width=1\textwidth]{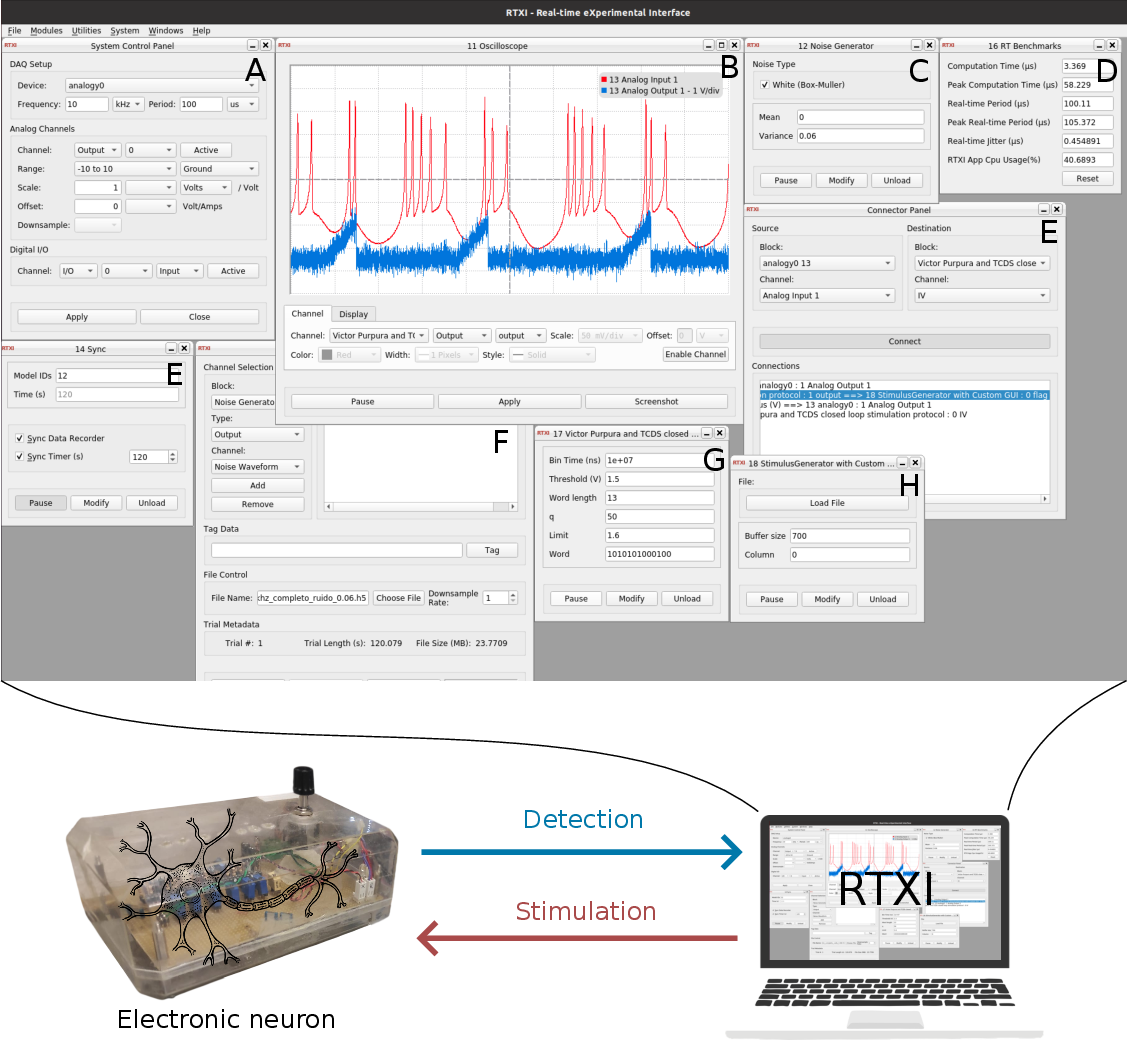}
	\caption{Hardware-constrained real-time setup. The activity produced by the Hindmarsh-Rose electronic neuron is acquired in real-time by the DAQ and processed by the modules implemented in the Real-Time eXperiment Interface (RTXI). The protocol module (B) analyzes the acquired signal and, when a code with a temporal structure similar to the triggering pattern is detected, the stimulus generator (C) delivers a stimulus to the system via the DAQ. The Noise Generator module (D) injects Gaussian noise to induce variability in the activity of the electronic neuron. Modules A, E, F, G, H, and I are used for experimental configuration, synchronization, data recording, analysis, and visualization.} \label{real-time-set-up}
\end{figure}

The real-time setup consists of both hardware and software components. The hardware included an electronic neuron that implemented the HR model, and a Data Acquisition Card (DAQ). As in the simulated environment, the electronic neuron served as the system to be conditioned, and its activity was acquired and digitized through the DAQ. The software component was developed in the Real-Time eXperiment Interface (RTXI) framework, a hard real-time data acquisition and control tool designed for biological experimentation. This modular platform enables real-time experiments using built-in and custom modules. For the experiments presented in this paper, modules that implement the protocol described in Section \ref{protocol-description} and the corresponding open‑loop protocol were developed as open-source software and made publicly available\footnote{Open-source code available: \href{https://github.com/GNB-UAM/Temporal-Code-Driven-Stimulation-and-Victor-Purpura-distance-protocol}{https://github.com/GNB-UAM/Temporal-Code-Driven-Stimulation-and-Victor-Purpura-distance-protocol}}. An example RTXI module configuration is shown in Figure \ref{real-time-set-up}. The following modules played a key role in the real-time experiments:

\textbf{System Control Panel}: Allows setting the experiment frequency and activating the input/output channels of the DAQ (marked as A in Figure \ref{real-time-set-up}). 

\textbf{Victor-Purpura and TCDS}: Processes the biological signal acquired through the DAQ and computes the Victor-Purpura distance between detected neural codes and the triggering code, as described in Section \ref{protocol-description}. If this distance falls below a predefined threshold, the module communicates with the Stimulus Generator module to triggering stimulation delivery (marked as B in Figure \ref{real-time-set-up}).

\textbf{Stimulus generator:} Delivers a pre-recorded stimulus to the system through an activated output channel of the DAQ (marked as C in Figure \ref{real-time-set-up}).

\textbf{Noise Generator:} Injects Gaussian noise into the analyzed system through a DAQ output channel, introducing variability in its activity (marked as D in Figure \ref{real-time-set-up}).

\textbf{Connector Panel:} Links the inputs and outputs among modules and DAQ channels (marked as E in Figure \ref{real-time-set-up}).

\textbf{RT Benchmarks:} Measures the real-time period and computation times of the tasks executed (marked as F in Figure \ref{real-time-set-up}). 

\textbf{Sync:} Synchronizes all modules involved in the experiment (marked as G in Figure \ref{real-time-set-up}).

\textbf{Oscilloscope:} Provides real-time visualization of DAQ inputs/outputs, module parameters, states, and other relevant information (marked as H in Figure \ref{real-time-set-up}).

\textbf{Data Recorder:} Records all module parameters, inputs, outputs, states, and DAQ channels throughout the experiments in real-time (marked as I in Figure \ref{real-time-set-up}).

\textbf{Additional modules:} Open-loop module, which delivers stimulation to the model at the same time points recorded during closed-loop stimulation. HR software model used in the regularization experiments. This model relies on a C++ library that implements multiple neural models \footnote{Open-source neural models library: \href{https://github.com/GNB-UAM/Neun}{https://github.com/GNB-UAM/Neun}}\cite{lareo2026neun}.

Furthermore, the specifications of the PC used to perform the experiments were as follows: Ubuntu 20.04, Xenomai 3.1, RTXI 2.4, and an Intel(R) Core(TM) i7-10700 processor.

\section{Results}

This section presents the characterization of the activity of the Hindmarsh-Rose neural model in closed-loop and open-loop sessions in both experiments, carried out in hardware-constrained and simulated environments. Section \ref{resBbs} presents the results obtained by conditioning the model activity to elicit brief bursts. Section \ref{resReg} shows how the model activity in a chaotic regime was regularized.

\subsection{Closed‑loop elicitation of brief bursts in the Hindmarsh–Rose model} \label{resBbs}

As explained in detail in Subsection \ref{ExpBbs}, closed-loop protocol adaptability was compared with an open-loop protocol. Gaussian noise was injected into the model to introduce variability in its activity.

In the simulated environment, the protocol’s ability to elicit brief bursts in the HR model was previously evaluated across a range of noise intensities from 0\% to 26\% \cite{ayala_2025}. Those results showed that, under closed-loop stimulation, the protocol could robustly trigger bursts even at moderate noise levels, whereas open-loop stimulation was much less reliable. These findings indicate that the closed-loop protocol adapts to the variability in the model’s activity, delivering stimuli at precise moments to consistently achieve the desired state. Building upon this foundation, the present work investigates whether the protocol maintains this performance in real-time experiments.

The protocol was then tested in the hardware-constrained real-time environment to generate brief bursts of spikes. Three noise levels were tested: no noise, 0.06, and 0.475 (standard deviation). The latter correspond to 3\% and 25\% noise levels in the simulated experiments. As Table \ref{table:bbs} shows, for closed‑loop stimulation with no injected noise, the percentage of brief burst generated is 97.27\%, with 0.06 noise, 97.45\% and with 0.475 noise, 68.56\%. For open-loop stimulation the results were 41.78\%, 38.24\% and 24.26\% respectively. These findings demonstrate that closed-loop stimulation generated brief bursts more robustly than open-loop stimulation by adapting to noise-induced variability in the neuron’s activity. These results validate those reported in Ayala et al. \cite{ayala_2025} in a realistic environment. 

\begin{figure}
	\centering
	\includegraphics[width=0.9\textwidth]{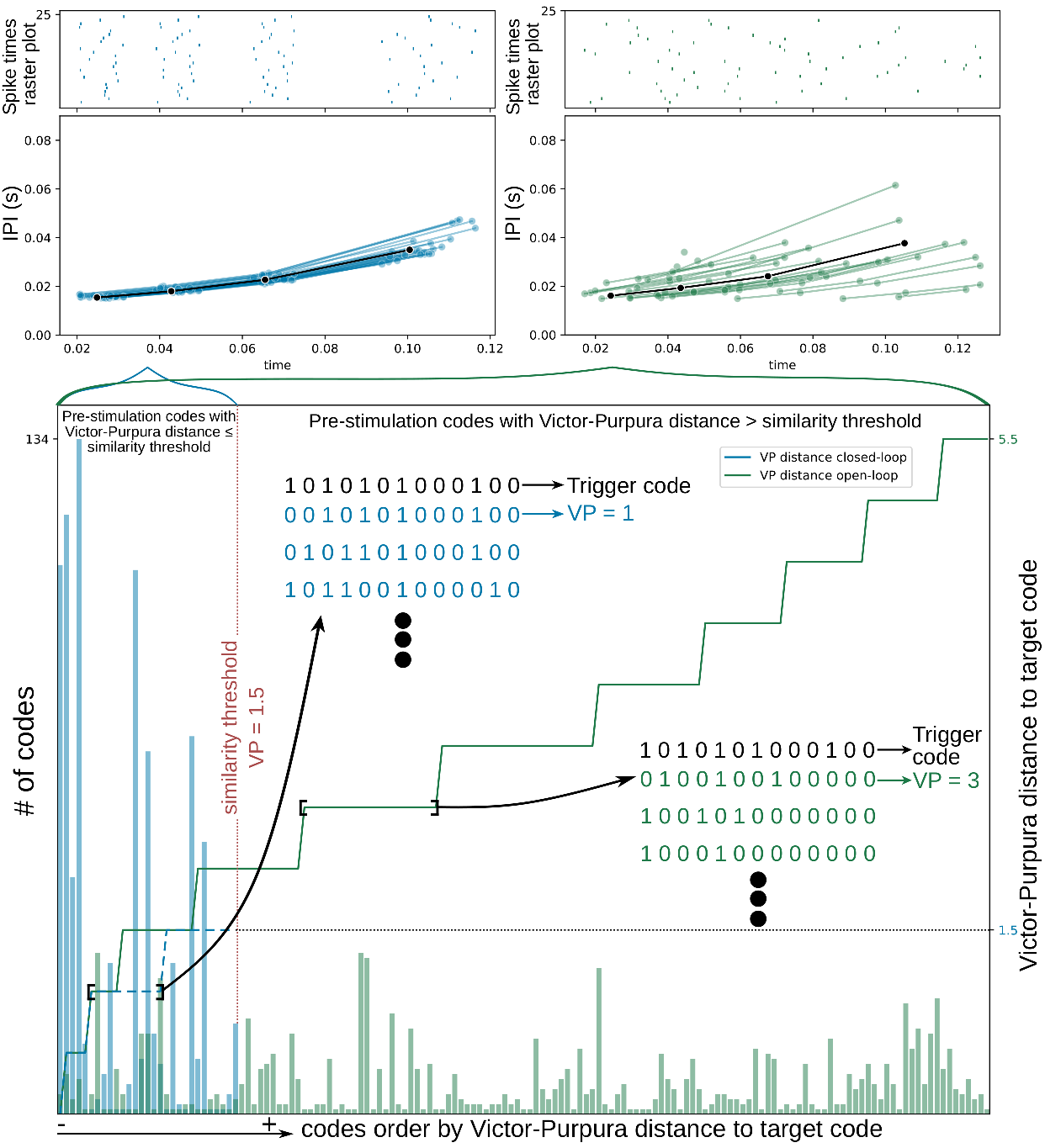}
	\caption{Victor-Purpura distance distribution computed between pre-stimulation spike codes and the triggering code in real-time experiments where brief bursts were generated. The histograms display the frequency of these Victor-Purpura distances. Closed-loop: Pre-stimulation codes exhibit a consistent temporal structure (Victor-Purpura distance below the threshold of 1.5), indicating effective burst detection and adaptation to variability. In the upper-left panel, the temporal structure of the pre-stimulation codes and the triggering code (plotted in black) is shown (inter-pulse intervals, IPIs), highlighting their similarity to the target pattern. Open-loop: Without adaptation (distances up to 5.5), stimulation often begins before burst completion, elongating bursts instead of eliciting brief ones. The IPIs of the pre-stimulation codes shown in the upper-right panel display a temporal structure that is not as close to the triggering code (plotted in black) as in the closed-loop sessions. In both upper panels, 25 pre-stimulation codes were selected randomly from the signals plotted in Figure \ref{Fig:bbs-codes-signal-rt}.
	} \label{Fig:bbs-codes-rt}
\end{figure}

\begin{figure}
	\centering
	\includegraphics[width=\textwidth]{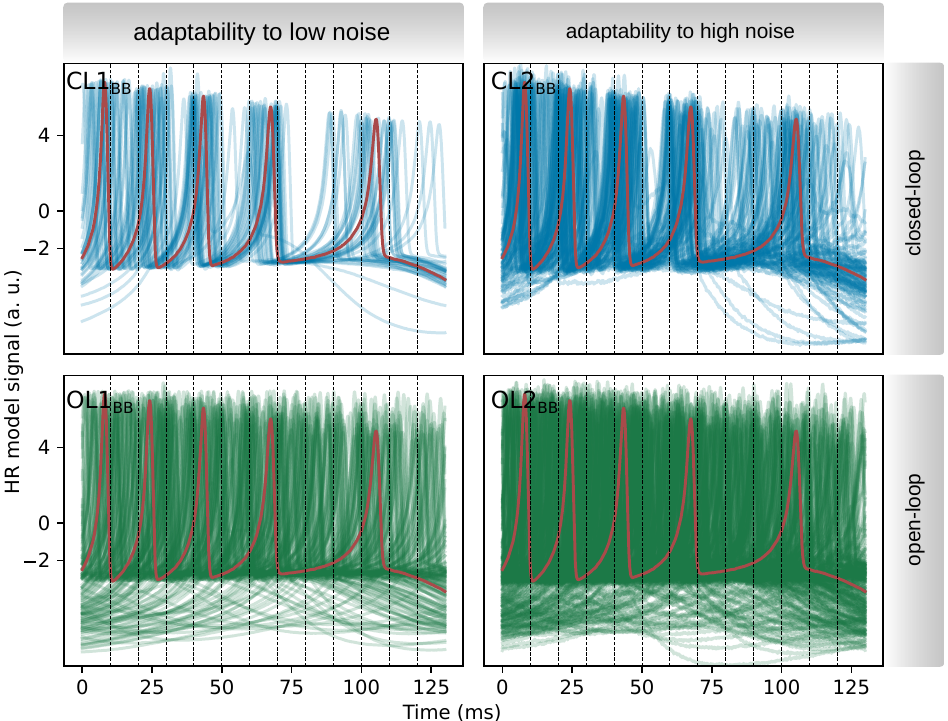}
	\caption{Record of the real-time HR model activity that triggers the stimulation in both closed-loop (blue) and open-loop (green) sessions. The target of these experiments was to elicit brief burst in the activity of the model. The adaptability of both methodologies to low noise introduced into the model is shown in panels labeled $CL1_{BB}$ (33 codes plotted) and $OL1_{BB}$ (149 codes plotted), corresponding to the Victor-Purpura distance distributions shown in Figure \ref{Fig:bbs-codes-rt}, the noise standard deviation was 0.06 (3\% noise level). The adaptability to high noise is shown in panels labeled $CL2_{BB}$ (99 codes plotted) and $OL2_{BB}$ (272 codes plotted), it was 0.475 (25\% noise level). The triggering is highlighted in red. The analysis of neural activity supports the findings shown in Figure \ref{Fig:bbs-codes-rt}. Furthermore, the differences between upper and bottom panels indicate that as the noise level injected into the HR model increases, the variability of the codes produced by the model also increases, reducing the performance of the closed-loop sessions.} \label{Fig:bbs-codes-signal-rt}
\end{figure}

Figure \ref{Fig:bbs-codes-rt} shows the Victor-Purpura distance between the spike sequences and the triggering code for closed-loop (blue) and open-loop (green) sessions (3\% noise, $\sigma = 0.06$). This figure illustrates that the pre-stimulation codes in the closed-loop session shared a similar temporal structure. In other words, the closed-loop protocol consistently detected spike sequences whose temporal structure resembled that of the triggering neural code (a complete burst). In contrast, pre-stimulation codes in open-loop sessions exhibited larger distance from the triggering code. Panels $CL1_{BB}$ and $OL1_{BB}$ in Figure \ref{Fig:bbs-codes-signal-rt} show the raw signals corresponding to the codes displayed in Figure \ref{Fig:bbs-codes-rt}. Panels $CL2_{BB}$ and $OL2_{BB}$ present the raw signals of the pre-stimulation codes for a 0.475 (25 \% noise level) noise standard deviation injected into the HR model. These observations demonstrate that the protocol can adapt to the intrinsic variability of the model, consistently detecting the triggering code. However, for pre-stimulation codes at a 0.475 noise standard deviation (panels $CL2_{BB}$ and $OL2_{BB}$), the diversity of detected codes increased, leading the protocol to identify some sequences that triggered stimulation but did not generate brief bursts.

\begin{figure}
	\centering
	\includegraphics[width=\textwidth]{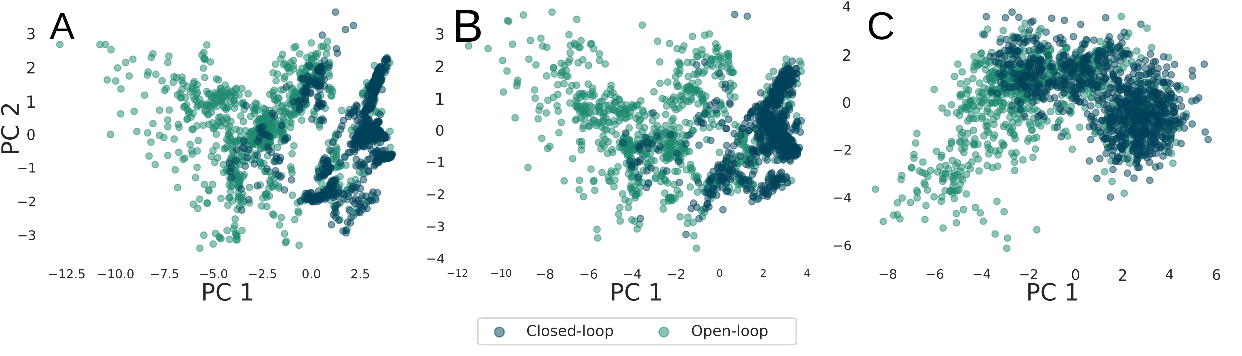}
	\caption{First two principal components of the PCA applied to the recurrence features extracted from the electronic neuron activity under closed-loop and open-loop stimulation. Panels A, B, and C show the data for no noise, 0.06 (3\% noise level), and 0.475 (25\% noise level) standard deviation noise, respectively. In each panel, blue dots correspond to the recurrence features extracted during closed-loop stimulation, and green dots to those extracted under open-loop stimulation. In Panel A, closed-loop data form a well-defined cluster, while green dots are more widely dispersed. This indicates that brief bursts were consistently generated in closed-loop sessions, in contrast to open-loop sessions. Panel B shows similar results to those in Panel A. However, the closed-loop data are less concentrated, indicating that brief bursts were generated less robustly. Panel C shows similar results to those in the other panels; however, the closed-loop points are more spread out. These findings indicate that as the noise level increased, closed-loop stimulation generated brief bursts less robustly, but still outperformed open-loop stimulation at each noise level. These results are consistent with the silhouette coefficients: 0.288  (Panel A), 0.274 (Panel B), 0.099 (Panel C).} \label{Fig:clustering-bbs-lab}
\end{figure}

Figure \ref{Fig:clustering-bbs-lab} shows the first two principal components from the PCA applied to recurrence features (analysis described in Section \ref{ExpBbs}) without noise (panel A), with noise of standard deviation 0.06 (panel B, 3 \% noise level), and with a standard deviation of 0.475 (panel C, 25 \% noise level) in the performed real-time experiments. In this analysis, each point represents a segment of activity characterized by recurrence metrics, allowing for a comparison of the dynamic structure of the activity generated under each condition. Panels A and B illustrate that the points corresponding to the closed-loop data (blue) formed a well-defined cluster, whereas those corresponding to the open-loop (green) data showed a scattered distribution. This indicates that brief bursts are consistently generated by the protocol, in contrast to open-loop stimulation. Similar results were obtained for 25\% noise (panel C). However, although the closed-loop points were clustered, they were more dispersed than in the previous closed-loop sessions, indicating that brief bursts are not generated with the same consistency. As noise increases, the effectiveness of the protocol decreased. This trend is reflected in the silhouette coefficients for increasing noise levels: 0.288 without noise, 0.274 for 3\% noise ($\sigma$=0.06), and  0.099 for 25\% noise ($\sigma$=0.475). These findings indicate that the protocol is able to condition the activity of the HR model to achieve the desired dynamic in a real-time environment adapting to the activity variability.

\begin{table}
	\centering
	\begin{tabular}{c c|c|c|c|} 
		\cline{3-5}
		&& \multicolumn{3}{c|}{Noise level} \\
		\cline{3-5}
		&&No noise&0.06 $\sigma$ (3\%)& 0.475 $\sigma$ (25\%)\\ [0.4ex] 
		\hline
		\multicolumn{1}{|c|}{\multirow{2}{*}{Experiment type}} &Closed-loop&97.27\%&97.45\% &68.56\%\\ 
		\cline{2-5}
		\multicolumn{1}{|c|}{}&Open-loop&41.78\%&38.24\%&24.26\%\\
		\hline
	\end{tabular}
	\caption{Percentage of brief bursts generated in real-time experiments. Six sessions (three trials) were performed for three different levels of noise injected into the Hindmarsh-Rose model in both closed-loop and open-loop conditions: no noise, 0.06, and 0.475 standard deviation.}
	\label{table:bbs}
\end{table}

\begin{table}
	\centering
	\resizebox{\textwidth}{!}{
	\begin{tabular}{c c|c|c|c|c|c|c|}
		\cline{3-8}
		&& \multicolumn{6}{c|}{Noise level} \\
		\cline{3-8}
		&& \multicolumn{2}{c|}{No noise} & \multicolumn{2}{c|}{0.06 $\sigma$ (3\%)} & \multicolumn{2}{c|}{0.475 $\sigma$ (25\%)} \\
		\cline{3-8}
		&\multicolumn{1}{c|}{} & Max (ns) & Avg (ns) & Max (ns) & Avg (ns) & Max (ns) & Avg (ns) \\
		\hline
		\multicolumn{1}{|c}{\multirow{3}{*}{Experiment type}}&\multicolumn{1}{|c|}{Control} &16937 &510.8& 6115 & 515.52& 7337 &  496.23\\
		\multicolumn{1}{|c}{}&\multicolumn{1}{|c|}{Closed-loop} & 10522 &555.16  &  7058& 515.65 & 8978 & 469.31 \\
		\multicolumn{1}{|c}{}&\multicolumn{1}{|c|}{Open-loop} &9398  & 474.98 &6777  &492.96  & 7471 & 583.38 \\
		\hline
	\end{tabular}}
	\caption{Average and maximum latency obtained in real-time experiments conditioning the activity of the Hindmarsh-Rose model to elicit brief bursts. Both average and maximum latency in all experiments are below the system period (10 kHz), meeting the temporal requirements for studying temporal coding in neural systems.}
	\label{table:lat-bbs-lab}
\end{table}

\begin{figure}
	\centering
	\includegraphics[width=\textwidth]{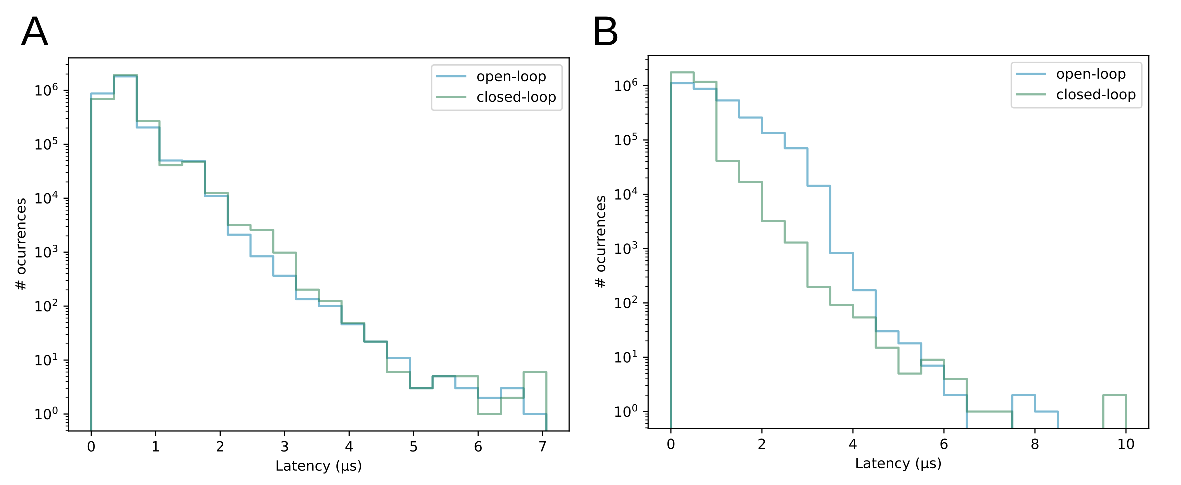}
	\caption{Latency distribution obtained in the real-time closed-loop and open-loop sessions for both experiments. Latency was calculated as the difference between the actual start time of tasks executed during an experimental period and the expected start time determined by the system period (100 \si{\micro\second}). Panel A shows latency distributions for the experiments in which brief burst were generated in the activity of the Hindmarsh-Rose model (0,06 noise standard deviation injected to the model). Panel B shows latency distributions for the experiments in which the chaotic activity of the HR model was regularized (0.015 noise standard deviation injected to the model). Latencies were below the system period.} \label{Fig:lat-exmaple}
\end{figure}

In addition, the protocol real-time capability was measured in these experiments. At a frequency of 10 kHz (sufficient for studying temporal coding), the results shown in Table \ref{table:lat-bbs-lab} were obtained. Across every experiment, both maximum and average latency remained below the system period (100 \si{\micro\second}), meeting the temporal requirements for studying temporal coding in these systems. As an example, the latency distribution of the closed-loop session with a noise standard deviation of 0.06 is shown in Panel A of Figure \ref{Fig:lat-exmaple}. These results demonstrate that closed-loop allows the stimulation timing to be adapted to the system's internal dynamics.

\subsection{Closed‑loop regularization of chaotic bursting in the Hindmarsh–Rose model}\label{resReg}

As introduced in Section \ref{ExpBbs}, the noise-induced chaotic activity of the
HR model was regularized using the closed-loop stimulation protocol and compared with an open-loop protocol.
\begin{figure}
	\centering
	\includegraphics[width=0.65\textwidth]{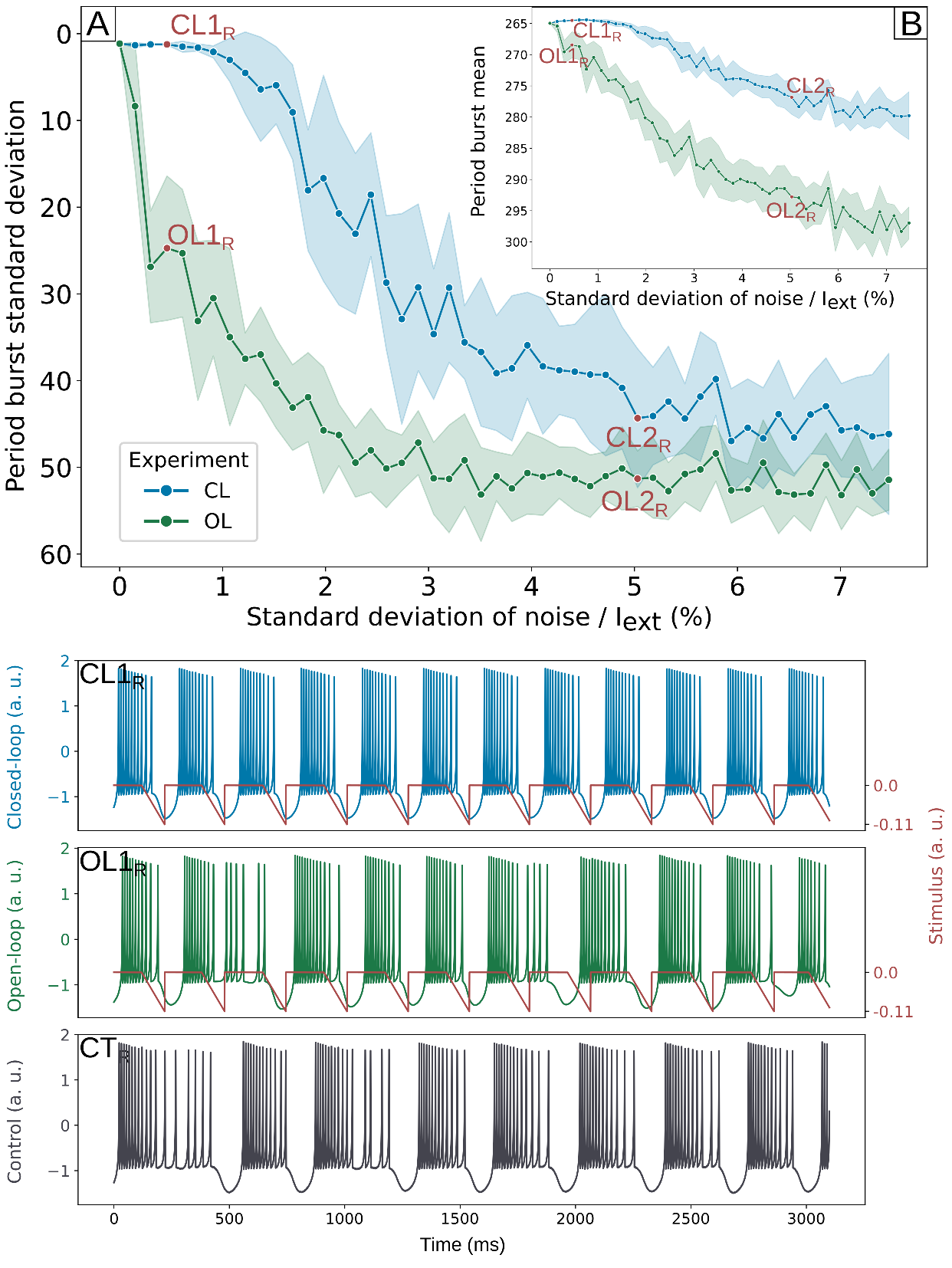}
	\caption{Analysis of the robustness with which the Hindmarsh-Rose model burst production is regularized while it is configured in a chaotic bursting regime, comparing the closed-loop stimulation protocol to the open-loop protocol in the simulated environment. In this analysis, Gaussian noise was incrementally supplied to the model as external current. Panel A shows the mean period standard deviation of bursts generated by the model for closed-loop and open-loop sessions as a function of the noise supplied. Panel B shows the mean period of bursts generated by the model for closed-loop and open-loop sessions as a function of the noise supplied. Panels $CL1_{R}$ and $OL1_{R}$ show an example of the activity of the HR model and the stimulus provided to it by the closed-loop and open-loop protocols, respectively, for a 0.45\% of noise. In closed-loop sessions, for lower noise values, the protocol drove the activity of the HR model to consistently regularize its burst production in contrast to open-loop experiments. For both methodologies, as the noise supplied to the model increases, the model generated less regular bursts. In addition, in closed-loop sessions, for lower noise values, the mean period remains constant. However, in the open-loop sessions, the mean period rises sharply as soon as noise was introduced to the model, indicating that the open-loop protocol conditioned the model in an less efficient way as the targeted goal.}\label{FigToleranciaCLOL3RegMeanStdPeriod}
\end{figure}

In closed-loop sessions, at low noise levels (noise < 1.5\%), the mean period deviation was close to zero, indicating that the protocol robustly regularized the HR model burst production (Figure \ref{FigToleranciaCLOL3RegMeanStdPeriod}). In contrast, in open-loop sessions, the mean period standard deviation increased significantly as soon as noise was introduced. In both cases, increasing noise led to higher variability, more pronounced under open-loop stimulation. These results indicate that the closed-loop protocol conditioned the HR model to produce regular bursts more robustly against noise-induced variability than the open-loop protocol, up to a 1.5\% noise.

Panel B shows the mean period of bursts produced by the HR model under closed-loop and open-loop stimulation. In closed-loop sessions, for noise < 2.5\%, the mean period remained constant. As the noise supplied to the model increased, the mean period gradually rose, although not abruptly. However, in open-loop sessions, the mean period increased sharply as soon as noise was introduced. These results indicate that the closed-loop regularized the activity while the open-loop did not. Panel $CL1_{R}$ provides an example of the HR model activity in a closed-loop session with 0.45\% noise, where burst production was consistently regularized. Panel $OL1_{R}$ shows another example of the model activity under the same conditions in an open-loop session, where regular bursts were not produced as robustly.

Additionally, this experiment was conducted in the hardware-constrained environment (as described in Section \ref{Exp-Reg}) for three different noise levels: no noise, 0.45\% noise ($\sigma = 0.015$), and 5\% noise ($\sigma = 0.165$). Table \ref{table:reg-lab-period} shows the results obtained for both closed-loop and open-loop conditions. When no noise was injected into the HR model, both methodologies achieved the same results, correctly regularizing its activity. This occurred because both experiments were performed under identical initial conditions, allowing the open-loop configuration to perfectly mimic the closed-loop regularization. However, when 0.45\% noise was introduced into the model, only the closed-loop stimulation robustly regularized the model activity, while the open-loop protocol failed to achieve this consistency. Finally, under 5\% noise, both methodologies failed to regularize the HR model activity. These findings are consistent with the results obtained in the simulated experiments shown in Figure \ref{FigToleranciaCLOL3RegMeanStdPeriod}.

\begin{figure}
	\centering
	\includegraphics[width=0.9\textwidth]{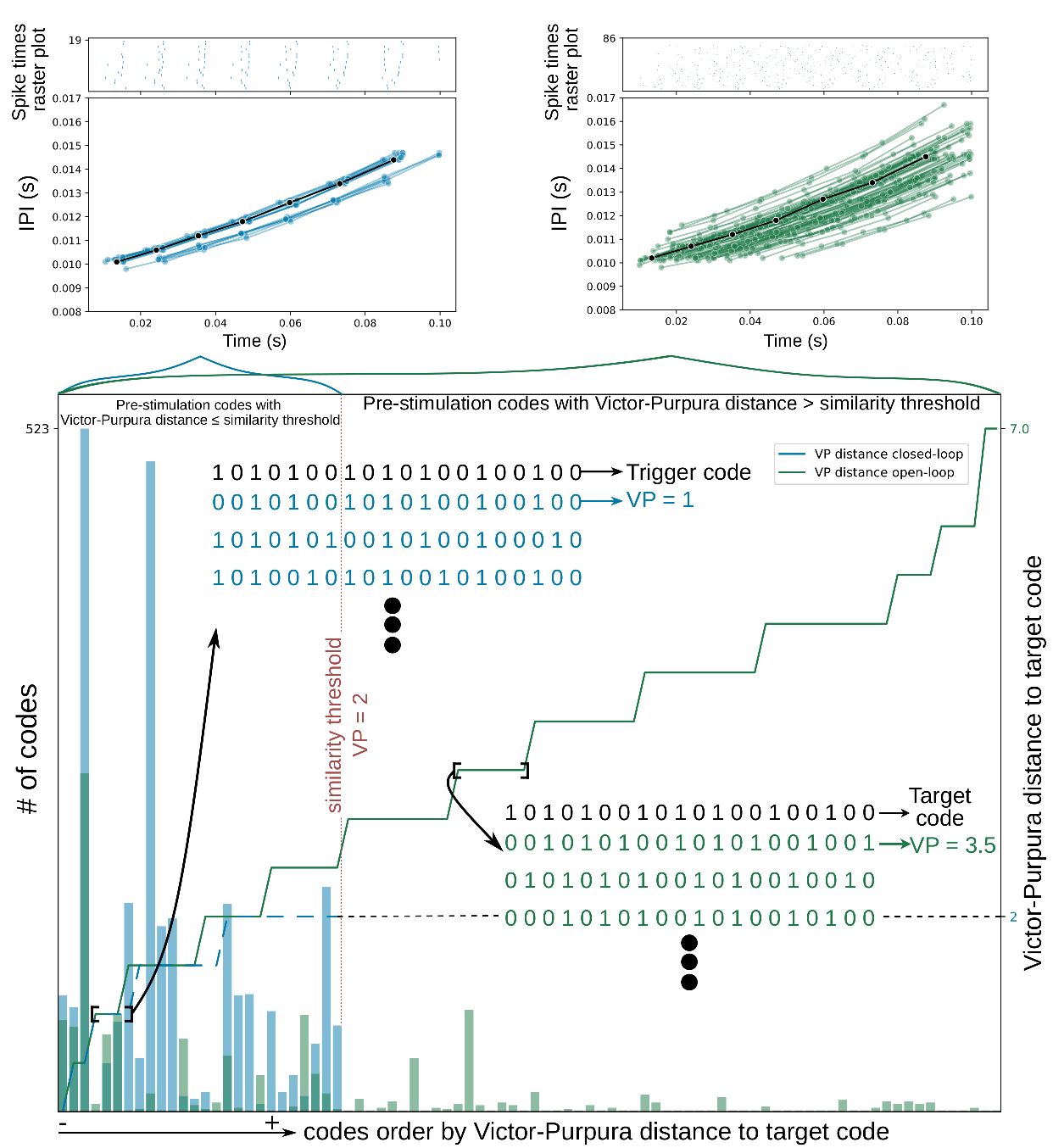}
	\caption{Victor-Purpura distance distribution computed between the neural codes produced by the Hindmarsh-Rose model immediately before stimulation and the triggering code in the real-time experiments where the model activity was regularized and  the noise level supplied to the model was 0.45\%. As in Figure \ref{Fig:bbs-codes-rt}, the histograms display the frequency of these Victor-Purpura distances. Closed-loop: Pre-stimulation codes exhibit a consistent temporal structure (Victor-Purpura distance below the threshold of 2), indicating effective burst detection and adaptation to variability. Upper-left panel shows a clear similarity between the temporal structure of the codes prior to stimulation and the triggering code (plotted in black). Open-loop: Failed to adapt to neural variability (Victor-Purpura distances reaching values up to 7). The IPIs of the pre-stimulation codes in the upper-right panel exhibit a structure that deviates more from the trigger (plotted in black) than in the closed-loop sessions.} \label{CodesBinReg}
\end{figure}

As in the brief-burst experiments, Figure \ref{CodesBinReg} shows that the Victor–Purpura distances of pre-stimulation codes in closed-loop (0.45\%) are close to the triggering code, in contrast to open-loop.

\begin{figure}
	\centering
	\includegraphics[width=0.83\textwidth]{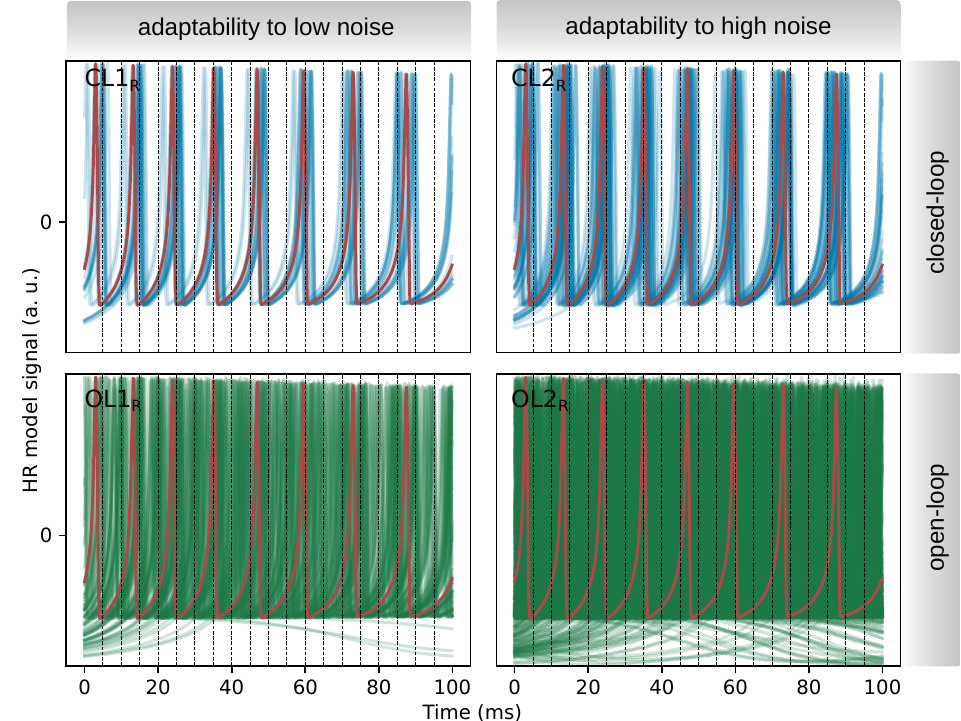}
	\caption{Illustration of the real-time HR model activity that triggers the stimulation in both closed-loop (blue) and open-loop (green) sessions. The target of these experiments was to regularize the activity of the chaotic Hindmarsh-Rose model. In panels $CL1_{R}$ (19 codes plotted) and $OL1_{R}$ (86 codes plotted), the noise level supplied to the model was 0.45\%, corresponding to the Victor-Purpura distribution shown in Figure \ref{CodesBinReg}; in panels $CL2_{R}$ (40 codes plotted) and $OL2_{R}$ (316 codes plotted), it was 5\% (see Figure \ref{FigToleranciaCLOL3RegMeanStdPeriod}). The code targeted by the protocol is highlighted in red. Each signal corresponds to a unique neural binary code, and all codes produced by the model are included. The raw signals of the pre-stimulation codes in closed-loop sessions exhibit a stereotyped structure similar to the triggering code, in contrast to the open-loop activity signals. These results demonstrate that the protocol was able to adapt to the variability of the neural code.} \label{Fig:codes-sig-reg-rt}
\end{figure}

Panels $CL1_{R}$ and $OL1_{R}$ in Figure \ref{Fig:codes-sig-reg-rt} show the raw signals of the pre-stimulation codes presented in Figure \ref{CodesBinReg} (0.45\%). Panels $CL2_{R}$ and $OL2_{R}$ present the same analysis for the experiments conducted at the 5\% noise level. In both cases, the closed-loop stimulation adapted to the variability with which the triggering code was emitted by the model, in contrast to the open-loop stimulation. In addition, as the noise level increased, greater variability was observed in the pre-stimulation codes. These results are coherent with those obtained in the brief burst generation experiments. However, as detailed in Table \ref{table:reg-lab-period} and illustrated in Figure \ref{CodesBinReg}, when 5\% noise was introduced into the model, the results of both closed-loop and open-loop stimulation sessions became similar, even though the protocol adapted to neural variability. This occurs because noise dominates system dynamics, overriding the effect of stimulation. Supporting evidence comes from the control experiments (results shown in Table \ref{table:reg-lab-period}), where, in the absence of stimulation, this level of noise caused the model activity to become more regular, effectively altering its baseline dynamic state.

\begin{figure}
	\centering
	\includegraphics[width=0.83\textwidth]{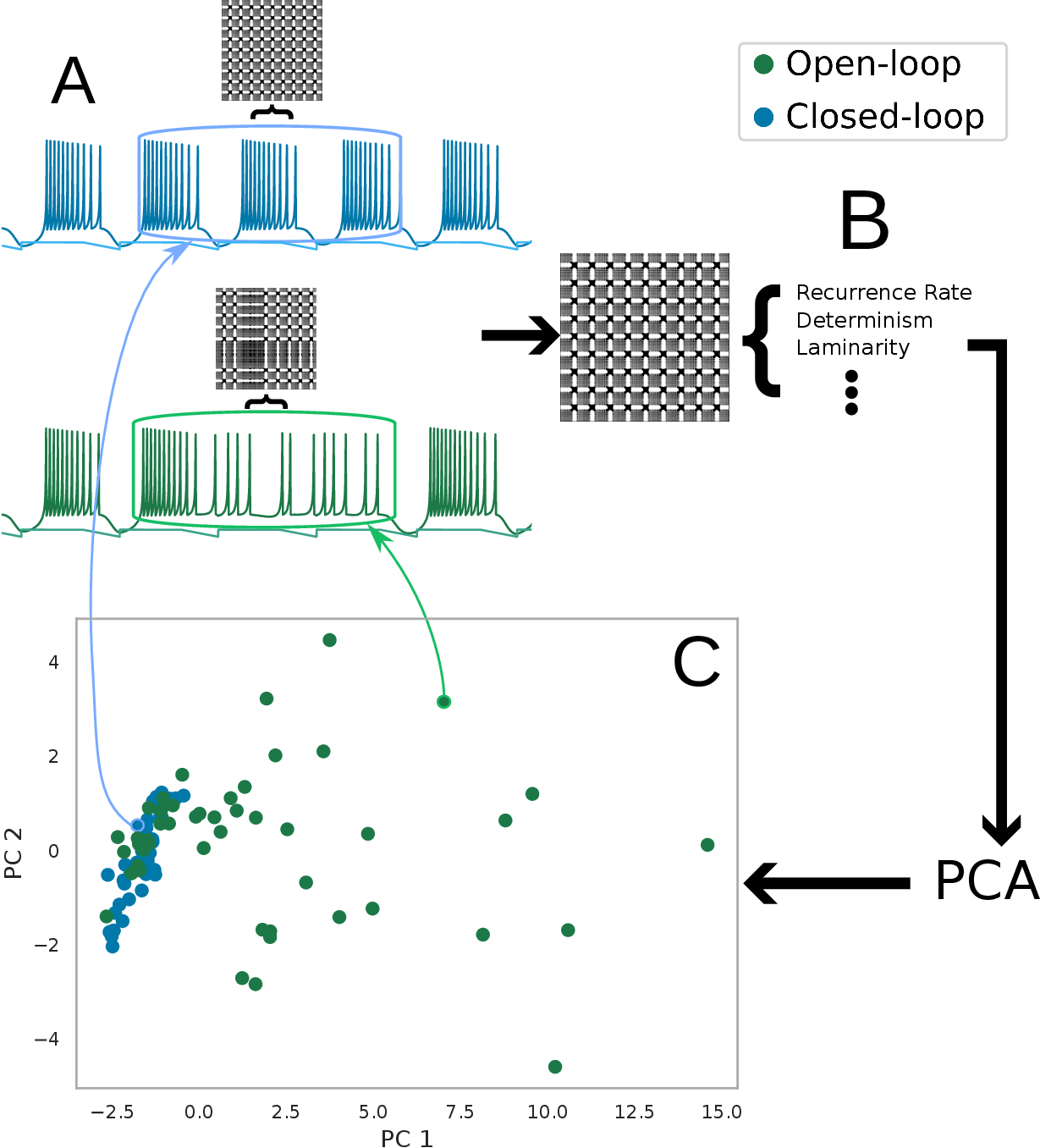}
	\caption{Flowchart of the recurrence plot computation, recurrence feature extraction, and PCA application in the real-time regularization of Hindmarsh-Rose model activity at a 0.45\% noise level. Panel A illustrates the random selection of signal traces in both closed-loop and open-loop sessions. For each selected signal segment, a recurrence plot was computed. From each recurrence plot, 15 metrics (laminarity, determinism, etc.) were extracted (Panel B). Finally, PCA was applied to these metrics, and the first two principal components were plotted (Panel C). In this experiment, closed-loop stimulation data exhibited a clustered structure in contrast to the scattered distribution observed in the open-loop condition. These findings indicate that closed-loop stimulation robustly regularized the Hindmarsh-Rose model activity, unlike the open-loop stimulation.} \label{Fig:cluster-reg-rt}
\end{figure}

Panel C of Figure \ref{Fig:cluster-reg-rt} shows the first two principal components obtained after applying PCA to the recurrence metrics extracted from random traces selected in closed-loop and open-loop sessions (as described in Section \ref{Exp-Reg}) in which 0.45\% noise was supplied to the model. The closed-loop data (blue points) form a well-defined cluster, in contrast to the open-loop data (green points), which are more dispersed. These findings indicate that closed-loop stimulation robustly regularizes the Hindmarsh-Rose model activity, unlike open-loop stimulation. Accordingly, the silhouette coefficient, equal to 0.163, indicates a clear difference between both methodologies.

Finally, Table \ref{table:lat-bbs-lab} details the real-time performance of these experiments at a frequency of 10 kHz (as in the brief burst experiment). Both the maximum and average latencies were below the system period, meeting the temporal requirements. As an example, the latency distribution of the closed-loop session with a noise level of 0.45\% is shown in Panel B of Figure \ref{Fig:lat-exmaple}.

\begin{table}
	\centering\resizebox{\textwidth}{!}{
		\begin{tabular}{c c|c|c|c|}
			\cline{3-5}
			&&\multicolumn{3}{c|}{Noise level}\\
			\cline{3-5}
			&& \multicolumn{1}{c|}{No noise} & \multicolumn{1}{c|}{0.015 $\sigma$ (0.45\%)} & \multicolumn{1}{c|}{0.165 $\sigma$ (5\%)} \\
			\cline{3-5}
			\hline
			
			\multicolumn{1}{|c}{\multirow{3}{*}{Experiment type}}&\multicolumn{1}{|c|}{Closed-loop} &$264.76\pm1.18$  &$264.48\pm1.21$  &$275.78\pm37.68$   \\
			\multicolumn{1}{|c}{}&\multicolumn{1}{|c|}{Open-loop} & $264.76\pm1.18$ & $267.55\pm20.73$&  $288.52\pm46.2$ \\
			\multicolumn{1}{|c}{}&\multicolumn{1}{|c|}{Control} &$556.57\pm162.24$& $462.87\pm180.1$ & $365.32\pm61.12$  \\
			\hline
	\end{tabular}}
	\caption{Mean and standard deviation of the period in the real-time activity regularization experiments. In the absence of noise, both closed-loop and open-loop stimulation correctly regularized the activity of the Hindmarsh-Rose model. When 0.45\% noise was injected, closed-loop stimulation robustly regularized its activity, in contrast to the open-loop condition. At a 5\% noise level, both methodologies failed to regularize the model activity.}
	\label{table:reg-lab-period}
\end{table}

\begin{table}
	\centering
	\resizebox{\textwidth}{!}{
	\begin{tabular}{c c|c|c|c|c|c|c|}
		\cline{3-8}
		&&\multicolumn{6}{c|}{Noise level}\\
		\cline{3-8}
		&& \multicolumn{2}{c|}{No noise} & \multicolumn{2}{c|}{0.015 $\sigma$ (0.45\%)} & \multicolumn{2}{c|}{0.165 $\sigma$ (5\%)} \\
		\cline{3-8}
		&\multicolumn{1}{c|}{} & Max (ns) & Avg (ns) & Max (ns) & Avg (ns) & Max (ns) & Avg (ns) \\
		\hline
		\multicolumn{1}{|c}{\multirow{3}{*}{Experiment type}}&\multicolumn{1}{|c|}{Control}&5670&232.47&8360&231.25  &6101& 279.42 \\
		\multicolumn{1}{|c}{}&\multicolumn{1}{|c|}{Closed-loop}&10344&850.6&10002&449.77   & 6616 &538.75 \\
		\multicolumn{1}{|c}{}&\multicolumn{1}{|c|}{Open-loop}&6648&446.27&8148&847.74  &8179& 486.38\\
		\hline
	\end{tabular}}
	\caption{Average and maximum latency obtained in real-time experiments regularizing the activity of the Hindmarsh-Rose. Both average and maximum latency in all experiments are below the system period (10 kHz), meeting the temporal requirements for studying neural coding in neural systems.}
	\label{table:lat-reg-lab}
\end{table}

\section{Discussion}

This work aimed to contribute to the field of real-time activity-dependent stimulation and, more specifically, to the study of temporal coding in neural systems. Closed-loop stimulation protocols have been developed and applied in many different contexts, e.g. \cite{Chamorro2012,sitaram_closed-loop_2017}. They have been widely used in the study of temporal information coding in sensorimotor processes \cite{muniz_real-time_2008}, in neuroethology \cite{Lareo2016,Lareo2017,Chamorro2012}, for monitoring \cite{sitaram_real-time_2011} and modulating brain activity \cite{zrenner_real-time_2018,ketz_closed-loop_2018,frohlich_closed-loop_2021}, as well as in the treatment of neurological disorders such as epilepsy \cite{patel_hard_2017}, schizophrenia \cite{frohlich_neuromodulation_2020}, and Parkinson’s disease \cite{little_adaptive_2013}, and in neurorehabilitation \cite{wolbrecht_optimizing_2008} or memory consolidation \cite{lustenberger_feedback-controlled_2016}, using both invasive and non-invasive stimulation. In this context, several protocols have been designed to detect the spatio-temporal activity of neural systems and drive it toward a desired dynamic state \cite{j_bryan_temporal_2025, wallach_neuronal_2011} while adapting stimulation parameters in real-time \cite{newman_optogenetic_2015, wallach_neuronal_2011}. The protocol studied here addresses closed‑loop stimulation from a temporal‑coding perspective, using real-time spike-train similarity metrics to guide system activity while explicitly accounting for spike-time variability. The ability to control neural systems activity using closed-loop protocols has been investigated in other preliminary studies, yielding similar findings \cite{newman_optogenetic_2015,wallach_neuronal_2011}, even when the same stimulus used in successful closed-loop sessions was applied in open-loop sessions.

Although the Hindmarsh-Rose model was used in this study as a case study due to its dynamic versatility, the protocol does not depend on any specific model. The detection–metric–stimulation chain remains valid as long as the system produces discretizable temporal patterns, making the approach general and applicable to biological systems. The decrease in performance at high noise levels reveals the limitations of similarity-based detection methods under conditions of extreme variability. 

The protocol studied in this work achieves adaptability to neural variability through the implementation of the Victor-Purpura distance. However, other metrics for comparing spike trains can be used for similar purposes. They can be classified into parametric and non-parametric measures. Among the most widely used parametric metrics are the Victor-Purpura distance \cite{Victor1996}, the van Rossum distance \cite{rossum_novel_2001}, and the Schreiber et al. \cite{schreiber_new_2003} similarity measure. These metrics are suitable for comparing complete spike trains and can be employed for the detection of stereotyped temporal patterns in real-time. Nevertheless, the Victor-Purpura distance is particularly intuitive, since its $q$ parameter directly defines the temporal scale of comparison. Moreover, the Victor-Purpura distance offers a direct interpretation in terms of spike operations (insertion, deletion, or displacement), which facilitates its tuning and implementation in online detection systems. In contrast, non-parametric measures such as the ISI-distance \cite{kreuz_measuring_2007}, SPIKE-distance \cite{kreuz_monitoring_2013}, and Event \cite{quian_quiroga_event_2002} and SPIKE \cite{kreuz_spiky_2015} synchronization are based on the local synchronization of firing times. These metrics are adaptive and do not require time-scale parameters. However, they present significant limitations for detecting stereotyped patterns: they only consider local spike-to-spike matches and fail to capture the sequential structure of the pattern or long-range temporal dependencies. Furthermore, they do not account for the total number of spikes, so spike trains with different firing rates can yield similar similarity values. 

It is important to highlight the advantages of the implementation in the RTXI software platform, rather than relying on hardware devices such as an FPGA. The protocol is standardized, easily disseminated, and user-friendly, allowing researchers to freely modify the parameters and even use other types of stimuli, offering great flexibility. The implementation of the protocol on other real-time experimental platforms will be explored to further facilitate its dissemination \cite{draelos_software_2025}.

\section{Conclusions}

A real‑time closed‑loop protocol guided by Victor–Purpura metrics robustly conditions Hindmarsh-Rose dynamics under variability (e.g., brief‑burst elicitation: 97.45\% vs 38.24\% closed‑ vs open‑loop; period deviation in regularization: $1.21$ ms vs $20.73$ ms). The closed-loop protocol studied in this work is a general-purpose tool for monitoring and activity-dependent stimulation of single-unit systems. The protocol detects sequences of spikes with a stereotypical temporal structure that may contain information about the underlying state of the system and triggers stimulation accordingly. This enables the study of whether the system activity leads to comparable responses. However, detecting these neural codes remains challenging due to the high variability with which they are produced.

To evaluate the protocol’s adaptability to neural variability, two experiments were designed using the Hindmarsh-Rose neural model as a case study: (i) generation of brief bursts and (ii) regularization of chaotic activity.In both cases, Gaussian noise was injected to induce variability. Performance was compared with an open-loop protocol in both simulated and real-time environments.

The results presented in Sections \ref{resBbs} and \ref{resReg} demonstrate that the protocol exhibits a high degree of adaptability to neural variability, successfully conditioning model activity in both methodologies. In contrast, open-loop stimulation failed to achieve the same robustness. These results confirm that adaptation based on the temporal structure of the code is key to the effective control of neural systems under variable conditions. However, as variability in the model activity increased, the effectiveness of the protocol declined. These findings suggest that implementing adaptive strategies to adjust the protocol parameters could further improve its performance, as other implementations have shown \cite{tafazoli_learning_2020,newman_optogenetic_2015, wallach_neuronal_2011}.

In addition, the observed latencies in the real-time environment meet the temporal requirements for studying fast neural dynamics, indicating that the protocol can be used to investigate temporal coding in a wide variety of biological systems.

These findings open opportunities for applying adaptive closed-loop protocols in experimental neuroscience and potentially in clinical contexts where variability is a major challenge. Real-time closed-loop stimulation enables robust adaptation to neural variability, allowing the study of temporal codes under realistic conditions. Future work will focus on applying the protocol in biological preparations to test whether variable spike trains detected as similar indeed constitute functionally equivalent neural codes. In addition, other similarity metrics, such as the van Rossum distance, and real-time machine learning approaches will be explored.

\subsubsection{Acknowledgments} This research was supported by grants PID2024-155923NB-I00, CPP2023-010818, PID2023-149669NB-I00 (MCIN/AEI and ERDF – “A way of making Europe”), and a grant from the Departamento de Ingeniería Informática at the Escuela Politécnica Superior of Universidad Autónoma de Madrid.

\bibliographystyle{elsarticle-num}
\bibliography{mybibliography}

\end{document}